\documentclass[preprint,prd,showkeys,showpacs,superscriptaddress,floatfix]{revtex4}

\usepackage{graphics,epsfig}
\usepackage{epstopdf}
\usepackage{subfigure}
\usepackage{palatino}
\usepackage[colorlinks=true,linkcolor=blue,urlcolor=blue,citecolor=blue]{hyperref}
\usepackage[toc,page]{appendix}
\usepackage[normalem]{ulem}
\usepackage{adjustbox}
\usepackage{latexsym}
\usepackage{amsmath}
\usepackage{amssymb}
\usepackage{amsfonts}
\usepackage{times}
\usepackage{graphicx}
\usepackage{dcolumn}
\usepackage{array}
\usepackage{wrapfig}
\usepackage{multirow}
\usepackage{tabularx}
\begin{document}

\title{Weak Deflection Angle and Greybody Bound of Magnetized Regular Black Hole}

\author{Wajiha Javed}
\email{wajiha.javed@ue.edu.pk} 
\affiliation{Department of Mathematics, Division of Science and Technology, University of Education, Lahore-54590, Pakistan}

\author{Sibgha Riaz}
\email{sibghariaz993@gmail.com} 
\affiliation{Department of Mathematics, Division of Science and Technology, University of Education, Lahore-54590, Pakistan}

\author{Ali {\"O}vg{\"u}n}
\email{ali.ovgun@emu.edu.tr}
\affiliation{Physics Department, Eastern Mediterranean University, Famagusta, 99628 North
Cyprus via Mersin 10, Turkey.}

\begin{abstract}In this paper, we examine the weak deflection angle and greybody bound for magnetized regular black hole.$~$ For this purpose, we apply the Gauss-Bonnet theorem on the black hole and obtain the deflection angle in plasma and non-plasma mediums.$~$ Moreover, we investigate graphically the effect of impact parameter on the deflection angle for regular black hole in both mediums. We examine that deflection angle goes to infinity when the impact parameter approaches to zero. We also observe that deflection angle shows negative behaviour at $q=0.6$ and $q=2.09$ but at $0.6<q<2.09$, angle shows the positive behaviour. Furthermore, we study the rigorous bound phenomenon of the greybody factor in the background for magnetized regular black hole. Later, we analyze the graphical behaviour of greybody bound with respect to different values of $\omega$ and observe that at small values of $\omega$, bound is increasing but for large values, bound is decreasing. Afterthat we examine that when we put $G=1$, $l=0$ and $q=0$, then all results for magnetized regular black hole solution reduces into results of the Schwarzschild black hole solution.

\end{abstract}

\pacs{95.30.Sf, 98.62.Sb, 97.60.Lf}
\keywords{General Relativity; Gravitational Lensing; Magnetized Black Holes; Gauss-Bonnet Theorem; Plasma Medium; Greybody Factory} 

\date{\today}
\maketitle

\section{Introduction}

Black Holes, a great prediction of Einstein's theory of General Relativity (GR) and at the same time the understandable objects inside the universe and are of most important for each classical and quantum gravity theories \cite{Einstein:1936llh}. A region of space having strong gravitational field that a matter or radiation, even a light, cannot escape from it is called Black Hole (BH). Initially BHs known as ``Collapser'', the term derived from the collapse of a star and later Wheeler put forward the term Black Hole \cite{Herdeiro:2018ldf}. A BH is an important tool for examining and testing the fundamental laws of the universe.$~$In fact, the Event Horizon Telescope collaboration captured the first image of a BH \cite{C1}.

In 1916, Einstein  anticipated the existence of gravitational lensing (GL) and gravitational waves as part of basics behind GR \cite{Einstein:1936llh}. Recently, the gravitational waves were detected by Laser Interferometer Gravitational-wave Observatory (LIGO) in 2015, which indicated that theoretical predictions are well expressed  with experimental observations \cite{LIGOScientific:2016emj}, After the detection of gravitational waves, a wide range of gravity theories faced many drawbacks, but the discovery of gravitational waves has gained interest in the field of GL \cite{Li:2018prc}. As the light emitted by distant galaxies passes by massive objects in the universe, the gravitational pull from these objects can distort or bend the light. This is called gravitational lensing. Gravitational lensing is a helpful method to understand the dark matter, galaxies and universe.

The GL has been categorized withinside the literature as strong GL, weak GL, micro GL \cite{Bartelmann:1999yn,Cunha:2018acu}. The method which is used to observe the magnification and position of a BH is known as strong GL. The first strong GL for the Schwarzschild BH was performed by Virbhadra and Ellis. Weak Gravitational lensing (WGL) is an effective tool to measure the masses of different objects in the universe.$~$Weak gravitational lensing investigates the cause of elevated enlargement of the universe and additionally distinguish among modified gravity and dark energy. Similarly, micro GL is worked on the bases of strong GL in which the image separation is too small to be resolved. Gravitational lensing were the processed in different space time in different ways \cite{Shaikh:2018lcc}-\cite{Kumar:2020hgm}.
Moreover, In past years, many studies have related GL with Gauss-Bonnet theorem (GBT) after Gibbons and Werner conformation about the useful method to calculate the bending angle of BHs that shows asymptotic behaviour \cite{Gibbons:2008rj}, which is given as:
\begin{equation}
\gamma=-\int \int_{D_\infty} \mathcal{K} dr,\nonumber\\
\end{equation}
where $\gamma$ represents the deflection angle, $\mathcal{K}$ represents the Gaussian optical curvature, $dr$ represents the optical surface and $D_\infty$ symbolizes the infinite domain of the space.

Werner extended GBT method to a stationary BH \cite{Werner:2012rc}. Ishihara et al \cite{Ishihara:2016vdc}, examined that it is possible to find a finite distance of deflection angle by using optical Fermat geometry. Crisnejo and Gallo \cite{Crisnejo:2018uyn} investigated the deflection angle of light in plasma medium. Since then, there is a constantly increasing interest to the WGL through the Gibbons and Werner technique using GBT methodology for BHs and wormholes \cite{Ishihara:2016sfv}-\cite{Javed:2020lsg}. In addition, Hensh et al. \cite{Hensh:2019ipu} computed the GL of Kehagias-Sfetsos compact objects surrounded by plasma.

In 1974, Hawking predicted that BHs emit quantum radiation. These radiation is known as Hawking radiation \cite{Hawking:1975vcx}. In the background of quantum field theory the creation and annihilation of the particles is theoretically possible. When pair production takes place close to the horizon of a BH, one of the particles present in the pair production falls and the other particle leaves the horizon of a BH, outside observer will detect this particle as Hawking radiation. However, according to GR, a BH bends spacetime around itself. This spacetime behaves like a gravitational potential for the particles to move. Few of the particles are reflected by the BH and remaining particles are transmitted through the BH.$~$As a result, the Hawking radiation that the observer observes from outside the BH is distinct from the propagation through the gravitational potential.$~$This distinction is known as the greybody factor. There are many studies investigating the computation of the greybody factor such as WKB approximation method \cite{Fernando:2004ay}-\cite{Konoplya:2020cbv}.$~$Another attractive ways to calculate the bound of greybody factor are \cite{Boonserm:2008qf}-\cite{Boonserm:2014fja}.

Singularity has been a major problem since the beginning of history \cite{Ca}. In GR, spacetime singularities increase several issues. Different techniques, such as nonlinear electrodynamics (NLED) and gravity modification, have been introduced to eliminate the spacetime singularities that occur at the centre of charged BHs. Born introduced the theory of NLED to solve point-like charge self-energy divergences and this theory was later modified with the help of Infield to establish the Born and Infield theory \cite{Cb}. NLED models have received a lot of attention in recent years, with a lot of focus on their ability to discover regular BH solutions.

A magnetic black hole is a black hole with a magnetic charge.~Magnetic black holes are permanent configurations compatible with the laws of physics. They have some interesting properties because they have large magnetic fields.~Electro-weak symmetry can be restored around the black hole due to magnetized black hole.~In astrophysics,~magnetized black holes have received a lot of attention because they're supposed to be good models for realistic stellar-mass and supermassive black holes.~Different methods have been proposed for obtaining electromagnetic energy from rotating magnetized black holes.~The motion of charged particles in the vicinity of magnetic black holes, and also its scattering and Hawking radiation, were all studied.~It was shown that in the presence of a magnetic field, rotating black holes' super-radiant instability and the intensity of Hawking evaporation are increased.~It has recently been proposed that certain particle collisions in the presence of a weakly magnetized non-rotating black hole can produce particles with high center-of-mass energy.~Under some instances, magnetized non-rotating black holes could be used as particle accelerators.~Exact solutions to the Einstein-Maxwell equations \cite{sibi1} can help us understand black hole astronomy.~The Harrison transformation was used to construct magnetized black hole solutions in $4$-dimensional spacetime. They've recently been extended to a number of Einstein-Maxwell and Einstein-Maxwell-dilaton equations describing black objects in external magnetic fields in $5$-dimensions.~We believe it is critical to obtain further magnetic solutions because only the simplest solution indicating a black hole on a gravitational instanton.~A mathematical formulation for a black hole which does not contain a singularity is known as Regular Black Hole.

The Main purpose of this paper is to analyze the magnetic regular black hole from global and analytical perspective in the presences of different mediums such as plasma and non-plasma by using Gauss-Bonnet theorem. Also, we calculate the bound of greybody factor for magnetized regular black hole. We examine the graphically impact of bound and graphical behaviour of plasma and non-plasma mediums on deflection angle.

This work is organized as, in section $\textbf{2}$, we study MRBH.$~$In section $\textbf{3}$, by using GBT, we calculate the deflection angle in the background of non-plasma medium.$~$In section $\textbf{4}$, we analyze the graphical behaviour of deflection angle in non-plasma medium.$~$In section $\textbf{5}$, we compute the deflection angle for MRBH in plasma medium and in section $\textbf{6}$ we study the graphical influence of deflection angle in plasma medium.$~$In the section $\textbf{7}$, we calculate the bound of greybody factor for MRBH. In the section $\textbf{8}$, we observe the graphical impact of greybody bound.$~$The last section $\textbf{9}$ is devoted to express the conclusion.

\section{ Magnetized Regular Black Hole (MRBH) }

A magnetic black hole is a black hole with a magnetic charge.~Magnetic black holes are permanent configurations compatible with the laws of physics.~They have some interesting properties because they have large magnetic fields.~Electro-weak symmetry can be restored around the black hole due to magnetized black hole.~In astrophysics, magnetized black holes have received a lot of attention because they're supposed to be good models for realistic stellar-mass and supermassive black holes.~A mathematical formulation for a black hole which does not contain a singularity is known as Regular Black.

Kruglov used the following Lagrangian of nonlinear electrodynamics to obtain the magnetized regular black hole \cite{Kruglov:2021mfy}

\begin{equation}
\mathcal{L}=-\frac{\mathcal{F}}{1+2 \beta \mathcal{F}}.
\end{equation}
with the field tensor:

\begin{equation}
\mathcal{F}=(1 / 4) F_{\mu \nu} F^{\mu \nu}= \left(\mathbf{B}^{2}-\mathbf{E}^{2}\right) / 2, F_{\mu \nu}=\partial_{\mu} A_{\nu}-\partial_{\nu} A_{\mu}.
\end{equation}

Note that the parameter $\beta$ of the dimension of (length) ${ }^{4}$, and there is a Maxwell limit $$
\mathcal{L} \rightarrow-\mathcal{F} .
$$ for weak fields $\beta \mathcal{F} \ll 1$. Then using the above nonlinear electrodynamics fields, Kruglov derived the MRBH metric in spherically coordinates as follows \cite{Kruglov:2021mfy},

\begin{equation}
 ds^2=-H(r)dt^2+ \frac{dr^2}{H(r)}+r^2d\Omega^2_{2},\label{IH1}
\end{equation}
where $H(r)$ can be defined as,
\begin{equation}
 H(r)=1-\frac{2Gm}{r}+\frac{Gq^2}{r^2}+\frac{4G^2m^2l^2}{r^4}-\frac{4G^2ml^2q^2}{r^5}+\mathcal{O}(r^{-6})\hspace{1.5em} r\rightarrow{\infty}, \label{SB1}
\end{equation}
while
\begin{equation}
d\Omega_{2} ^{2} = d\theta^2+\sin^2\theta d\phi^2. \label{SB2}
\end{equation}
Here $G$ is a Newton's constant, $l$ is the fundamental length and $m$ is the mass which is constant and $q$ is the magnetic charge. For equatorial plane $\theta=\frac{\pi}{2}$  and null geodesic $ds^{2}$=0, the optical metric is written as:
 \begin{equation}
dt^2=\frac{dr^2}{H(r)^2}+\frac{r^2d\phi^2}{H(r)}. \label{Cp7}
 \end{equation}
The non-zero Christoffel symbols of Eq.(\ref{Cp7}) calculates as:
\begin{equation}
\Gamma^0_{00}=-\frac{H^\prime(r)}{H(r)},~~~~~~~~~\nonumber\\
\end{equation}

\begin{equation}
\Gamma^1_{10}=\frac{1}{r}-\frac{H^\prime(r)}{2H(r)},~~~~~\nonumber\\
\end{equation}

\begin{equation}
\Gamma^0_{11}=\frac{-2H(r)}{r}+{H^\prime(r)}, \label{Cp8}
\end{equation}
in which $0$ and $1$ indicate $r$-coordinate and $\phi$-coordinate and the Ricci scalar of the optical metric computes as:
\begin{equation}
\mathcal{R}=H(r){H^{\prime\prime}(r)}-\frac{(H^{\prime}(r))^2}{2}. \label{Cp9}
\end{equation}
The Gaussian curvature can be defined as given in below expression;
\begin{equation}
 \mathcal {K}=\frac{Ricci~Scalar}{2}. \label{SB3}
\end{equation}
The required Gaussian optical curvature for the optical metric for MRBH is obtained as following;
 \begin{eqnarray}
\mathcal{K}&\approx&\frac{(3q^2-2mr)G}{r^4}+\frac{(-150mq^2+100m^2r+2q^4r-6mq^2r^2+3m^2r^3)G^2}{r^7}\nonumber
\\&-&\frac{2(47mq^4+9l^2mq^4-150m^2q^2r-15l^2m^2q^2r+78m^3r^2+6l^2m^3r^2)G^3}{r^9}\nonumber
\\&-&\frac{2(485m^2q^4-360l^2m^2q^4-780m^3q^2r +600l^2m^3q^2r+6mq^6r)G^4}{r^{12}}\nonumber
\\&-&\frac{2(-3l^2mq^6r+312m^4r^2-240l^2m^4r^2-18m^2q^4r^2+9l^2m^2q^4r^2)G^4}{r^{12}}\nonumber
\\&-&\frac{2(18m^3q^2r^3-9l^2m^3q^2r^3-6m^4r^4+3l^2m^4r^4)G^4}{r^{12}}.
\end{eqnarray}

\section{ Weak Deflection Angle for MRBH in Non-Plasma Medium}

According to GBT, the topology of domain $(\mathcal{S}_{R})$ and the intrinsic structural of the spacetime are connected
with boundary ${\partial\mathcal {S}}_{R}$. Then, with the help of GBT, the angle of deflection of the desired BH can be calculated as, \cite{Gibbons:2008rj}
\begin{equation}
  \int\int_{\mathcal{S}_{R}}\mathcal{K}dS+\oint_{\partial\mathcal{S}_{R}}kdt
 +\sum_{i}\epsilon_{i}=2\pi\mathcal{X}(\mathcal{S}_{R}), \label{SB4}
\end{equation}
the geodesic curvature $k$ revealed as $k=\bar{g}(\nabla_{\dot{\beta}}\dot{\beta},\ddot{\beta})$ such that $\bar{g} (\dot{\beta},\dot{\beta})=1$. If  ${\beta}$ is consider to be a smooth curve then we obtain $\dot{\beta}$ as a unit speed vector and $\epsilon_{i}$ is represents the exterior angle at the $ith$ vertex. As $R\rightarrow\infty$, then we take $\pi/2$ as jump angle.
This gives us ($\theta_{ \mathcal{S}}+\theta_{\mathcal{O}}\rightarrow\pi$).
\begin{equation}
 \int\int_{\mathcal{S}_{R}}\mathcal{K}dS+\oint_{\partial
 \mathcal{S}_{R}}kdt+\epsilon_{i}=2\pi\mathcal{X}(\mathcal{S}_{R}), \label{SB5}
\end{equation}
Here, the total jump angle is presents by $\epsilon_{i}=\pi$. $\mathcal{X}$ is representing Euler characteristic number which is $1$. Geodesic curvature
takes the form $k(C_{R})=\mid\nabla_{\dot{P}_{R}}\dot{P}_{R}\mid$ that we will compute.~The radial part mention in the geodesic curvature calculates as:
\begin{equation}
 (\nabla_{\dot{P}_{R}}\dot{P}_{R})^{r}=\dot{P}^{\phi}_{R}
 \partial_{\phi}\dot{P}^{r}_{R}+\beta^{r}_{\phi\phi}(\dot{P}^{\phi}_{R})^{2}.\label{IQ4}
\end{equation}
For high value of $R$, $P_{R}:=r(\psi)=R=const,$ where $R$ shows the distance from the coordinate origin. Last expression $\beta^{r}_{\psi\psi}$ shows the Christoffel symbols in connection to the optical geometry. Due to the non-presence of the topological effect the first term present in the above expression vanishes while the second term takes the form $k(P_{R})=\mid\nabla_{\dot{P}_{R}}\dot{P}_{R}\mid$ that shall be computed by using of unit speed term.
\begin{equation}
 (\nabla_{\dot{P}^{r}_{R}}\dot{P}^{r}_{R})^{r}\rightarrow\frac{1}{R}.\label{SB6}
\end{equation}
$k(P_{R})\rightarrow R^{-1}$ due to absences of topological effect. By taking the advantage of the optical metric;
it can be written $dt=Rd\phi$. It can be stated that:
\begin{equation}
 k(P_{R})dt=\frac{1}{R}Rd\phi.\label{SB7}
\end{equation}
The following expression can be written when we collect all the above results;
\begin{equation}
 \int\int_{\mathcal{S}_{R}}\mathcal{K}ds+\oint_{\partial \mathcal{S}_{R}} kdt
 \overset{R \rightarrow\infty }{=}\int\int_{O_{\infty}}\mathcal{K}dS+\int^{\phi+\gamma}_{0}d\phi.\label{iq2}
\end{equation}
The photon ray is expressed as  $r(t)=b/\sin\phi$ at $0th$ order in weak field deflection limit. Therefore the angle of deflection is defined as: \cite{Gibbons:2008rj}
\begin{equation}
 \gamma=-\int^{\pi}_{0}\int^{\infty}_{b/\sin\phi}\mathcal{K}\sqrt{det\bar{g}}drd\phi,\label{IQ5}
\end{equation}
 $\sqrt{det\bar{g}}$ is calculated as:
\begin{equation}
\sqrt{det\bar{g}}=r(1-\frac{2Gm}{r}+\frac{Gq^2}{r^2}+\frac{4G^2m^2l^2}{r^4}-\frac{4G^2ml^2q^2}{r^5})~~~~~~~~~~
\end{equation}
By exploit the above relation and putting the values of Gaussian curvature into Eq.(\ref{IQ5}), angle of deflection $\gamma$ can be written as;
\begin{eqnarray}
  \gamma&\approx&\frac{(4Gm)}{b}+\frac{(3G^{2}m^{2}\pi)}{4b^{2}}-\frac{(3Gq^{2}\pi)}{4b^{2}}-\frac{(8G^{2}mq^{2})}{3b^{3}}-\frac{(75G^{2}m^{2}\pi)}{8b^{4}}\nonumber
\\&-&\frac{(45G^{3}m^{2}q^{2}\pi)}{32b^{4}}+\frac{(15G^{2}q^{4}\pi)}{64b^{4}}+\frac{(32G^{2}mq^{2})}{b^{5}}+\frac{(12G^{3}mq^{4})}{5b^{5}}\nonumber
\\&+&\frac{(125m^{2}G^{3}q^{2}\pi)}{8b^{6}}+\frac{(525G^{4}m^2q^{4}\pi)}{256b^{6}}-\frac{(4192G^{3}mq^{4})}{245b^{7}}\nonumber
\\&-&\frac{(43155G^{4}m^{2}q^{4}\pi)}{2048b^{8}}+\frac{(6111G^{4}m^{2}q^{4}\pi)}{256b^{10}}+\frac{(567G^{4}l^{2}m^{2}q^{4}\pi)}{128b^{10}}.\label{S1}
\end{eqnarray}

The deflection angle ${\gamma}$ obtain in non-plasma medium depends on the $m$ of the BH, $q$, $b$, $G$ and $l$. We  observe that in the attained deflection angle the first term is the well known result for the Schwarzschild black hole, when we put $G=1$, $l=0$, $q=0$.

\section{Graphical Representation for Non-Plasma of MRBH }

This section is for the graphical impact of the deflection angle $\gamma$ on MRBH. The physical importance of the graphs is too studied in order to analyze the effect of different parameters on the achieved angle by allocating different values to BH charge, impact parameters and fundamental length.
\begin{center}
\epsfig{file=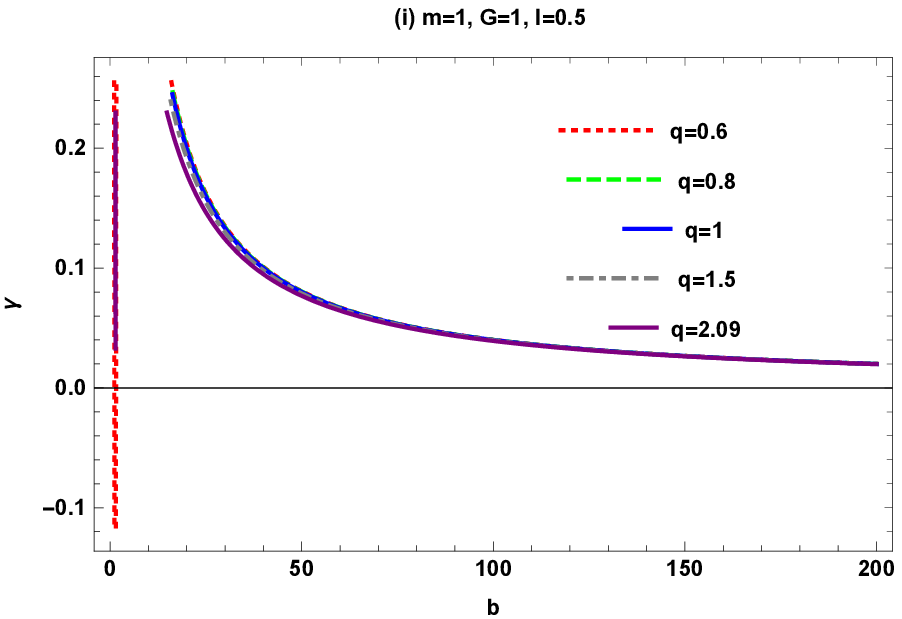,width=0.50\linewidth}\epsfig{file=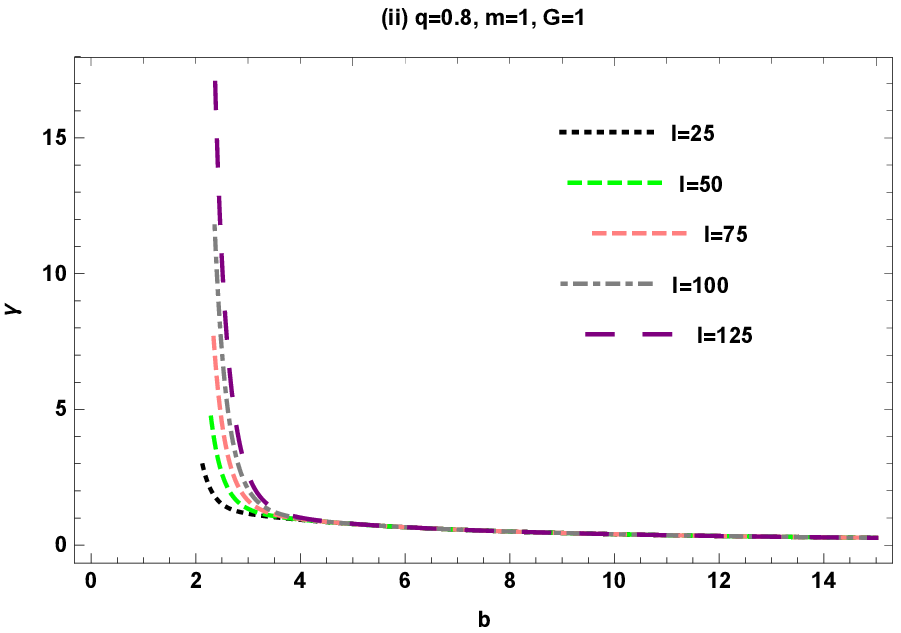,width=0.50\linewidth}\\
{Figure 1: $\gamma$ in connection with $b$}.
\end{center}
\subsection{Angle of Deflection $\gamma$ versus Impact Parameter $b$}

\textbf{Figure 1} shows the conduct of $\gamma$ with respect to $b$ by taking the fix value of $m=1$, $G=1$ and varying $q$ and $l$, respectively.

In the first figure, it is observed that deflection angle shows the negative behaviour at $q=0.6$ and $q=2.09$ and shows positive behaviour at $0.6<q<2.09$ by fixing $G=1$, $m=1$ and $l=0.5$.

In the second figure, we analyzed that the deflection angle gradually decreasing and then eventually goes to infinity for large variation of $l$ and fix $G=1$, $m=1$ and $q=0.8$, which is the unstable state of magnetized BH. Therefore, we conclude that for small values of $l$ the MRBH is stable but as $l$ increases it shows the unstable behavior of MRBH. We also observe that at high value of $m$ the angle shows the same behaviour.

\begin{center}
\epsfig{file=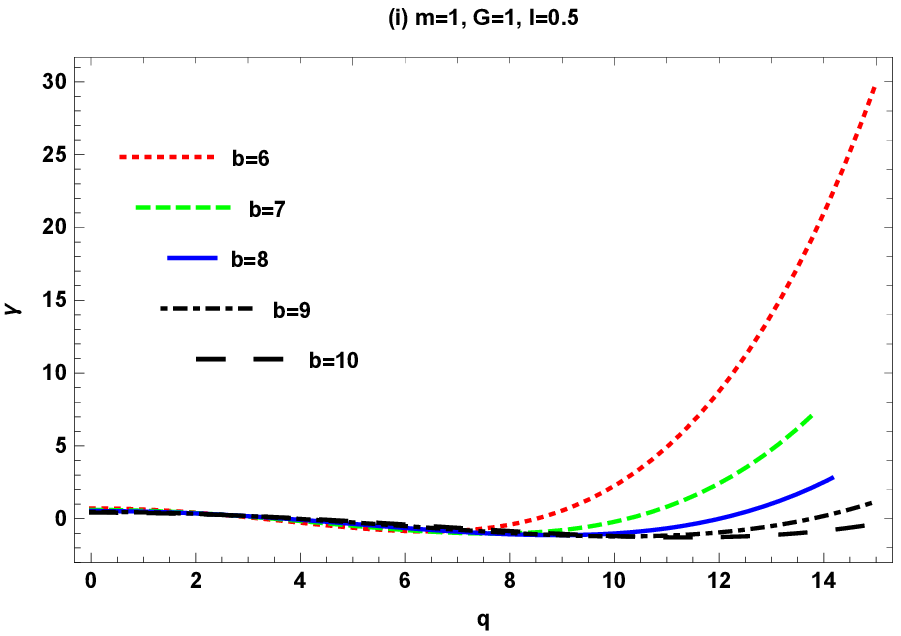,width=0.50\linewidth}\epsfig{file=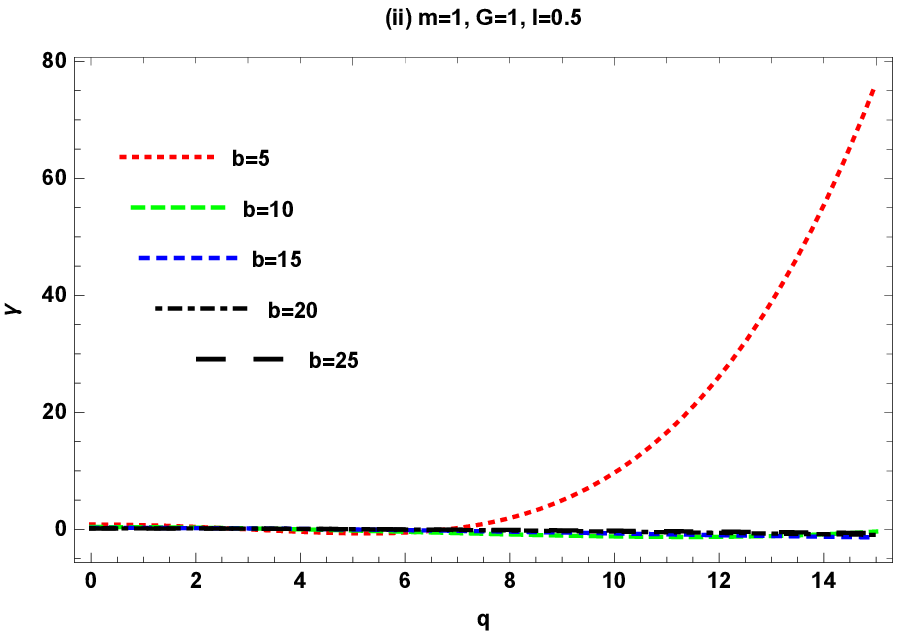,width=0.50\linewidth}\\
{Figure 2: $\gamma$ in connection with $q$}.
\end{center}

\subsection{Deflection Angle $\gamma$ versus BH Charge $q$}

\textbf{Figure 2} shows the behaviour of $\gamma$ with $q$ by fixing $l=0.5$, $G=1$, $m=1$ and varying $b$. The domain of BH charge is taken to be $0\leq{q}\leq 15$.

In the left graph, it is observe that deflection angle firstly increase but then for higher values of $b$ the deflection angle is decreased. We also observe that when we fix greater value of $l$, the angle shows the same behaviour.

In the right graph, We observed that the deflection angle is increasing for smaller values of b but as $b$ increases the deflection angle become negatively decreased. Also, we examined that angle shows the same behaviour when we put $m=150$.

\section{Effect of Plasma on Weak Deflection Angle $\gamma$ for MRBH}

In this section the impact of the angle of deflection by MRBH in plasma medium is examine. We consider the MRBH in plasma possessing refractive index, \cite{Crisnejo:2018uyn}
\begin{equation}
n^2\left(r\right)=1-\frac{\omega_e^2(r)}{\omega_\infty^2(r)}H(r). \label{SB7}
\end{equation}
The corresponding refractive index is express as:
 \begin{equation}
n(r)=\sqrt{1-\frac{\omega_e^2}{\omega_\infty^2}{H(r)}}.\label{SB8}
 \end{equation}
Here, $\omega_e$ denotes the electron plasma frequency and $\omega_\infty$ denotes photon frequency.
The corresponding metric function is written as:
\begin{equation}
 ds^2=-H(r)dt^2+ \frac{1}{H(r)}dr^2+r^2d\Omega_{2}^2, \label{SB9}
\end{equation}
and
\begin{equation}
 H(r)=1-\frac{2Gm}{r}+\frac{Gq^2}{r^2}+\frac{4G^2m^2l^2}{r^4}-\frac{4G^2ml^2q^2}{r^5}.\label{SB10}
\end{equation}
In order to study the application of the Gauss-Bonnet theorem to the determination of the bending angle proposed by Gibbons and Werner \cite{Gibbons:2008rj}, we consider a two dimensional Riemannian manifold $(M^{opt}; g^{opt}_ {xy})$ with the optical metric $g^{opt}_ {xy}=-\frac{n^2}{g_{00}}g_{xy}$. The corresponding optical metric is define as following: \cite{Crisnejo:2018uyn}
\begin{equation}
 d\sigma^2=g^{opt}_{xy}dx^xdx^y=n^2 \left[\frac{dr^2}{H^2(r)}+\frac{r^2d\vartheta^2}{H(r)}\right],\nonumber
\end{equation}
\begin{equation}
where~~~~~~~~x,y=1,2,3,......... \label{iq3}~~~~~~~~~~~~~~~~~~~~~~~~~~~~~~\nonumber
\end{equation}
This metric preserves the angle between two curves at a given point and is conformally related to the metric \ref{IH1}, when source and observer are in the tropical region possess $\theta=\frac{\pi}{2}$ and we imposed $ds^{2}$=0 in case of null geodesics.

Gaussian optical curvature $\mathcal{K}$ is examine as;
\begin{equation}
    \mathcal{K}=\frac{R_{r\psi r\psi}(g^{opt}_{xy})}{det(g^{opt}_{xy})}.\label{iq1}
\end{equation}
we let for simplicity:
\begin{eqnarray}
\tilde{\omega}&=& \frac{\omega_e}{\omega_\infty}
\end{eqnarray}
By using Eq.(\ref{iq1}) in the weak field approximation Gaussian curvature is express as:
 \begin{eqnarray}
\mathcal{K}&\approx& (-\frac{2}{r^3}-\frac{3\tilde{\omega}^2}{r^3})mG+(\frac{3}{r^4}+\frac{5\tilde{\omega}^2}{r^4})q^2G+\frac{100+3r^2}{r^6}m^2G^2\nonumber \\&+&\frac{12(15+r^2)\tilde{\omega}^2}{r^6}m^2G^2+(\frac{-6(25+r^2)}{r^7}-\frac{(275+26r^2)\tilde{\omega}^2}{r^7})mq^2G^2\nonumber
\\&+&(\frac{2}{r^6}+\frac{(10){\tilde{\omega}}^2}{r^6})q^4G^2-\frac{12(13+l^2)}{r^7}m^3G^3+\frac{30(10+l^2)}{r^8}m^2q^2G^3\nonumber
\\&-&\frac{2(362+13l^2+6r^2){\tilde{\omega}}^2}{r^7}m^3G^3+\frac{2(737+34l^2+16r^2){\tilde{\omega}}^2}{r^8}m^2q^2G^3\nonumber
\\&-&\frac{2(47+9l^2)}{r^9}mq^4G^3-\frac{(560+42l^2+23r^2){\tilde{\omega}}^2}{r^9})mq^4G^3\nonumber
\\&+&\frac{6(260-200l^2-6r^2+3l^2r^2)}{r^{11}}m^3q^2G^4-\frac{2(1212){\tilde{\omega}}^2}{r^{11}}m^3q^2G^4\nonumber 
\\&-&\frac{2(1592l^2){\tilde{\omega}}^2}{r^{11}}m^3q^2G^4-\frac{2(941r^2+75l^2r^2){\tilde{\omega}}^2}{r^{11}}m^3q^2G^4\nonumber 
\\&-&\frac{2(485-360l^2-18r^2+9l^2r^2)}{r^{12}}m^2q^4G^4+\frac{4(370){\tilde{\omega}}^2}{r^{12}}m^2q^4G^4\nonumber
\\&+&\frac{4(480l^2+336r^2+35l^2r^2){\tilde{\omega}}^2}{r^{12}}m^2q^4G^4.
\end{eqnarray}
By using GBT deflection angle is calculate and contrast to the equation of angle obtain for non-plasma. Since light beams follows a straight line approximation and calculate the deflection angle for plasma.
\begin{equation}
    \gamma=-\lim_{R\rightarrow 0}\int_{0} ^{\pi} \int_\frac{b}{\sin\psi} ^{R} \mathcal{K} dS \label{sib}.
\end{equation}
By using Eq.(\ref{sib}), we obtain the angle of the desired BH for plasma medium expresses as;
\begin{eqnarray}
  \gamma&\approx&\frac{(4Gm)}{b}+\frac{(3G^{2}m^{2}\pi)}{4b^{2}}-\frac{(3Gq^{2}\pi)}{4b^{2}}-\frac{(8G^{2}mq^{2})}{3b^{3}}-\frac{(75G^{2}m^{2}\pi)}{8b^{4}}\nonumber
\\&-&\frac{(45G^{3}m^{2}q^{2}\pi)}{32b^{4}}+\frac{(15G^{2}q^{4}\pi)}{64b^{4}}+\frac{(32G^{2}mq^{2})}{b^{5}}+\frac{(12G^{3}mq^{4})}{5b^{5}}\nonumber
\\&+&\frac{(125m^{2}G^{3}q^{2}\pi)}{8b^{6}}+\frac{(525G^{4}m^2q^{4}\pi)}{256b^{6}}-\frac{(4192G^{3}mq^{4})}{245b^{7}}\nonumber
\\&-&\frac{(43155G^{4}m^{2}q^{4}\pi)}{2048b^{8}}+\frac{(6111G^{4}m^{2}q^{4}\pi)}{256b^{10}}+\frac{(567G^{4}l^{2}m^{2}q^{4}\pi)}{128b^{10}}\nonumber
\\&+&\frac{(2Gm){\tilde{\omega}}^2}{b}-\frac{(15G^2m^2\pi){\tilde{\omega}}^2}{2b^4}-\frac{(G^2m^2\pi){\tilde{\omega}}^2}{2b^2}+\frac{(80G^2mq^2){\tilde{\omega}}^2}{3b^5}\nonumber
\\&+&\frac{(2G^2mq^2){\tilde{\omega}}^2}{b^3}-\frac{(Gq^2\pi){\tilde{\omega}}^2}{2b^2}+\frac{(945G^4l^2m^2q^4\pi){\tilde{\omega}}^2}{256b^{10}}\nonumber
\\&-&\frac{(465G^3m^2q^2\pi){\tilde{\omega}}^2}{32b^6}+\frac{(3G^3m^2q^2\pi){\tilde{\omega}}^2}{8b^4}+\frac{(4112G^3mq^4){\tilde{\omega}}^2}{245b^7}\nonumber
\\&-&\frac{(2G^3mq^4)\omega^2}{3b^5}-\frac{(3G^2q^4\pi){\tilde{\omega}}^2}{16b^4}-\frac{(15435G^4m^2q^4\pi){\tilde{\omega}}^2}{256b^{10}}\nonumber
\\&+&+\frac{(14105G^4m^2q^4\pi){\tilde{\omega}}^2}{2048b^8}-\frac{(45G^4m^2q^4\pi){\tilde{\omega}}^2}{128b^6}.\label{S2}
\end{eqnarray}
In case of plasma medium, the deflection angle ${\gamma}$ depends on the $m$ of the BH, $q$, $b$, $G$, $l$ and on the plasma term. The bending angle obtain in the plasma medium increases with the parameter $\frac{\omega_e^2}{\omega_\infty^2}$, which shows that lower the photon frequency observe by a static spectator  at infinity, greater the deflection angle of it for the fix electron plasma frequency. We also observe that when we take the $q=0$, $G=1$, $l=0$, the deflection angle obtain in the plasma medium reduces to the deflection angle of the Schwarzschild black hole in plasma medium. We also investigate that the deflection angle obtain in the plasma medium reduces to the deflection angle that we have obtained in case of non-plasma, when we take $\frac{\omega_e^2}{\omega_\infty^2}={\tilde{\omega}}=0$.

\section{Graphical Analysis for Plasma Medium}

This section is to study the effect of plasma on deflection angle $\gamma$. The physical importance of these figure are studied. Moreover, we consider  $\tilde{\omega}=10^{-1}$ and give different values to the $b$, $q$ of the BH, to investigate the behaviour of $\gamma$.
\begin{center}
\epsfig{file=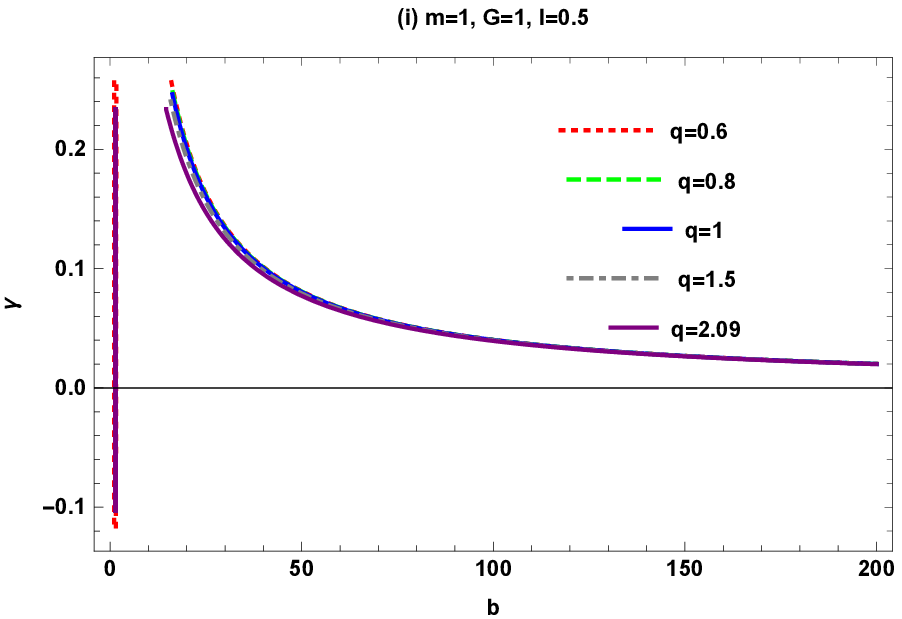,width=0.50\linewidth}\epsfig{file=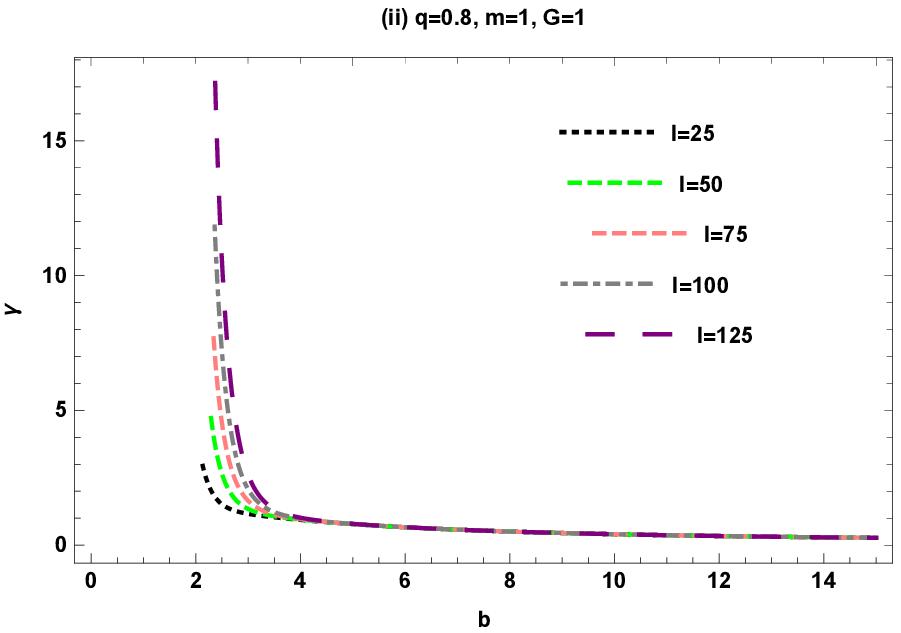,width=0.50\linewidth}\\
{Figure 1: $\gamma$ in connection with $b$}.
\end{center}

\subsection{Deflection Angle $\gamma$ with Impact Parameter $b$}

In the first figure, it is observed that deflection angle shows the negative behaviour at $q=0.6$ and $q=2.09$ and shows positive behaviour at $0.6<q<2.09$ by fixing $G=1$, $m=1$ and $l=0.5$.

In the second figure, we analyzed that the deflection angle gradually decreasing and then eventually goes to infinity for large variation of $l$ and fix $G=1$, $m=1$ and $q=0.8$, which is the unstable state of magnetized BH. Therefore, we conclude that for small values of $l$ the MRBH is stable but as $l$ increases it shows the unstable behavior of MRBH. We also observe that at high value of $m$ the angle shows the same behaviour.

\begin{center}
\epsfig{file=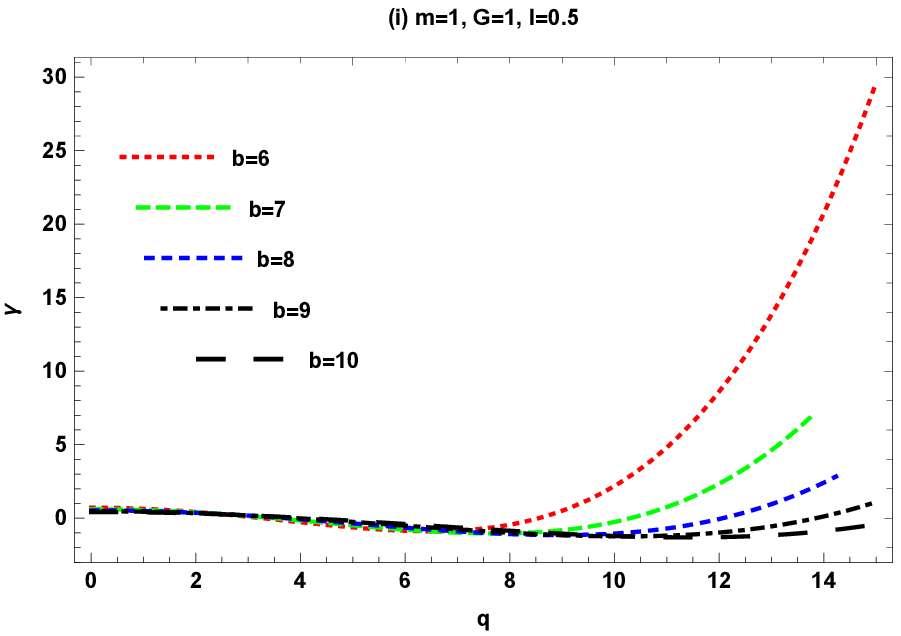,width=0.50\linewidth}\epsfig{file=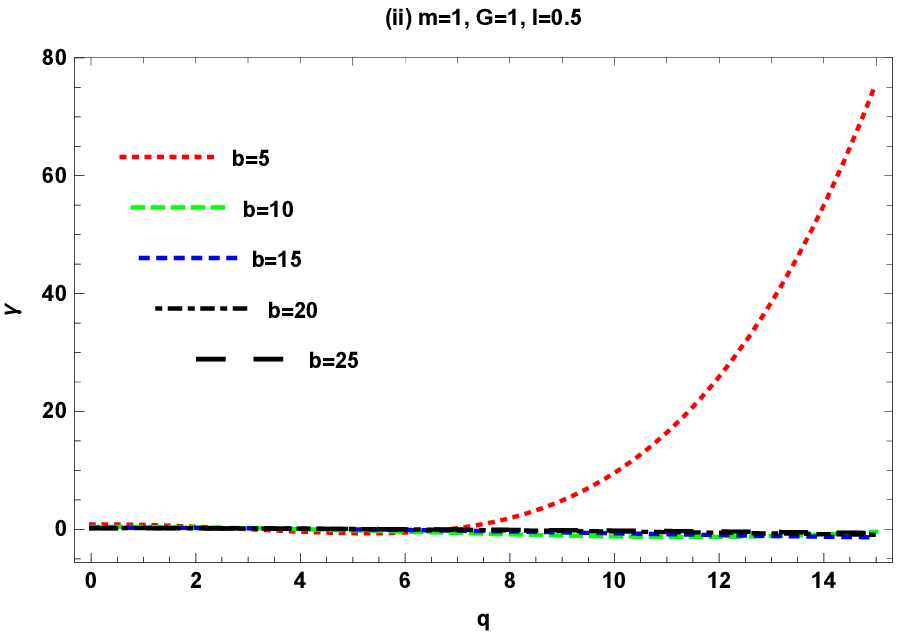,width=0.50\linewidth}\\
{Figure 2: $\gamma$ in connection with $q$}.
\end{center}

\subsection{Deflection Angle $\gamma$ with BH Charge $q$}

In the left graph, it is observe that deflection angle firstly increase but then for higher values of $b$ the deflection angle is decreased. We also observe that when we fix greater value of $l$, the angle shows the same behaviour.

In the right graph, We observed that the deflection angle is increasing for smaller values of b but as $b$ increases the deflection angle become negatively decreased. Also, we examined that angle shows the same behaviour when we put $m=150$.

\section{Derivation of Greybody Factor of MRBH}

The MRBH metric in d-dimension is defined as,
\begin{equation}
 ds^2=-H(r)dt^2+ \frac{dr^2}{H(r)}+r^2 (d\theta^2+\sin^2\theta d\phi^2),\label{SB11}
\end{equation}
where the  metric function is defined as:
\begin{equation}
 H(r)={1-\frac{2Gm}{r}+\frac{Gq^2}{r^2}+\frac{4G^2m^2l^2}{r^4}-\frac{4G^2ml^2q^2}{r^5}}.\label{SB12}
\end{equation}
We computed numerically the horizons of the metric $H(r)$.
\begin{center}
\epsfig{file=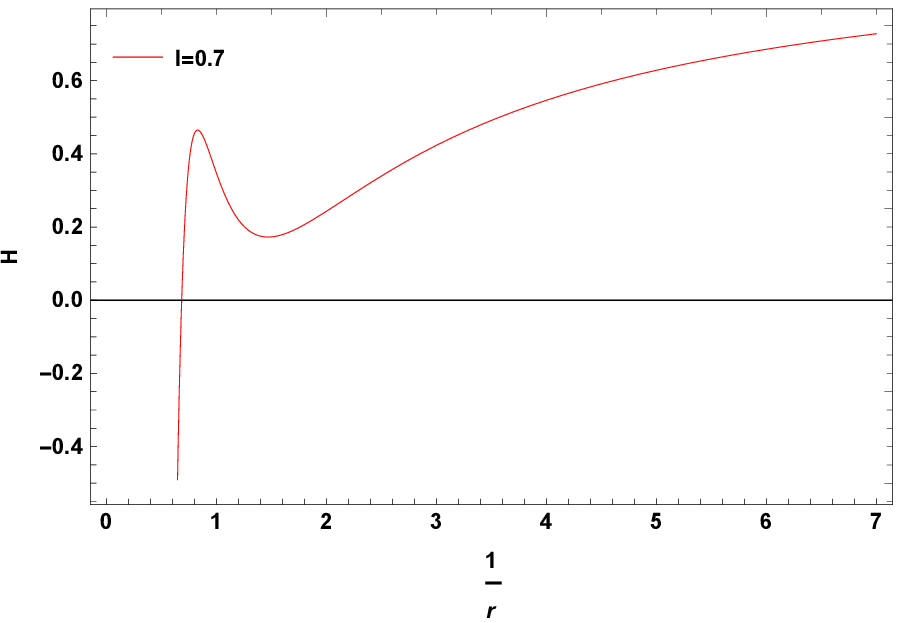,width=0.50\linewidth}\\
{Figure 1: $H$ versus $r$}.
\end{center}
 After the computation, the outer horizon of MRBH is ${r_+}=1.6$. Schrodinger like equation are defined as: \cite{Boonserm:2008qf,Boonserm:2009zba,Boonserm:2008zg}
\begin{equation}
\left[\dfrac{d^2}{dr_*^2}+\omega^2-{\text{V}}(r)\right]\phi=0,\label{SB13}
\end{equation}
where
\begin{equation}
dr_*=\frac{1}{H(r)}dr,\label{SB14}
\end{equation}
The potential $\text{V}(r)$ mention in the Eq.(\ref{SB13}) is stated as:
\begin{equation}
{V}(r)=\frac{(d-2)(d-4)}{4}\frac{H^2(r)}{r^2}+\frac{(d-2)}{2}\frac{H(r)\partial_rH(r)}{r}+l(l+d-3)\frac{H(r)}{r^2}.\label{SB15}
\end{equation}
Greybody factor bound is calculated as using the following equation \cite{Boonserm:2008zg}
\begin{equation}
    T\geq sech^2\left(\frac{l}{2\omega}\int_{r_+} ^{\infty}\frac{{V}(r)}{H(r)}dr_*\right).\label{SB16}
\end{equation}
Then
\begin{equation}
    T\geq sech^2\left(\frac{1}{2\omega}\int_{r_+} ^{\infty}(\frac{(d-2)(d-4)}{4}\frac{H(r)}{r^2}+\frac{(d-2)}{2}\frac{\partial_rH(r)}{r}+\frac{l(l+d-3)}{r^2})dr\right)\label{SB17}
\end{equation}
The Eq.(\ref{SB17}) is written as when we put d=4,
\begin{equation}
    T\geq sech^2\left(\frac{1}{2\omega}\int_{r_+} ^{\infty}(\frac{\partial_rH(r)}{r}+\frac{l(l+1)}{r^2})dr\right)\label{SB18}
\end{equation}
After simplifying the integral, we substituted ${r_+}$ ,
\begin{equation}
T\geq sech[\frac{0.625l(l+1)+0.390625Gm-0.305176{G}^2{l}^2{m}^2-0.16276{G}{q}^2+0.198682{G}^2{l}^2{m}{q}^2}{2\omega}]^2. \label{SB19}
\end{equation}

Hence we obtain the final expression for rigorous bound of the MRBH and observe that it depends on $l$, $G$, $q$ and $m$ and $\omega$. \\

\section{Graphical Study of the Greybody Bound for MRBH }

In this section we analyze the graphical behaviour of greybody bound on MRBH. The physical significance of these plots is also shows the effect of parameters on the lower bound by changing the values of the charge of the BH.

\subsection{Rigorous bound $T_b$  in connection with omega $\omega$}

\begin{center}
\epsfig{file=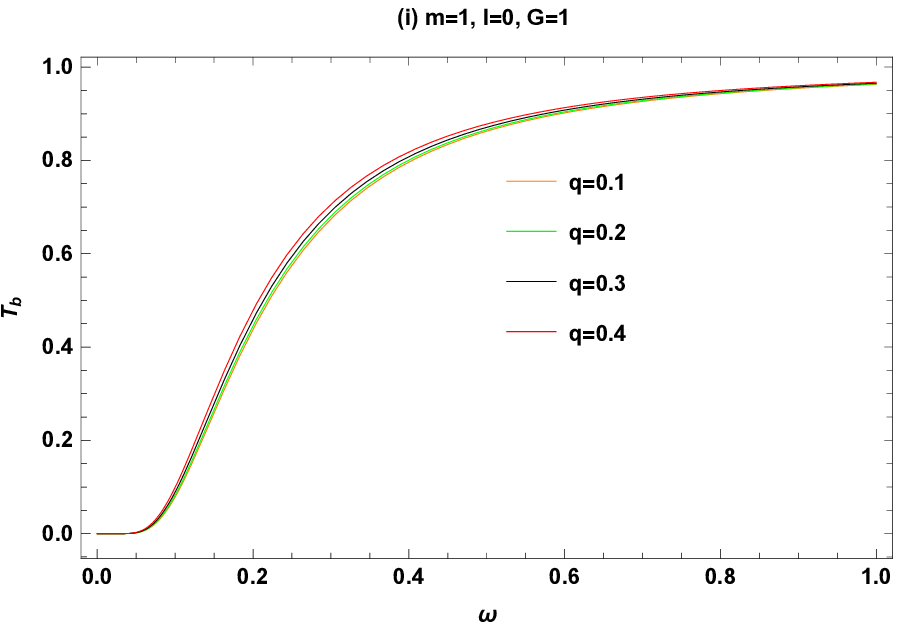,width=0.50\linewidth}\epsfig{file=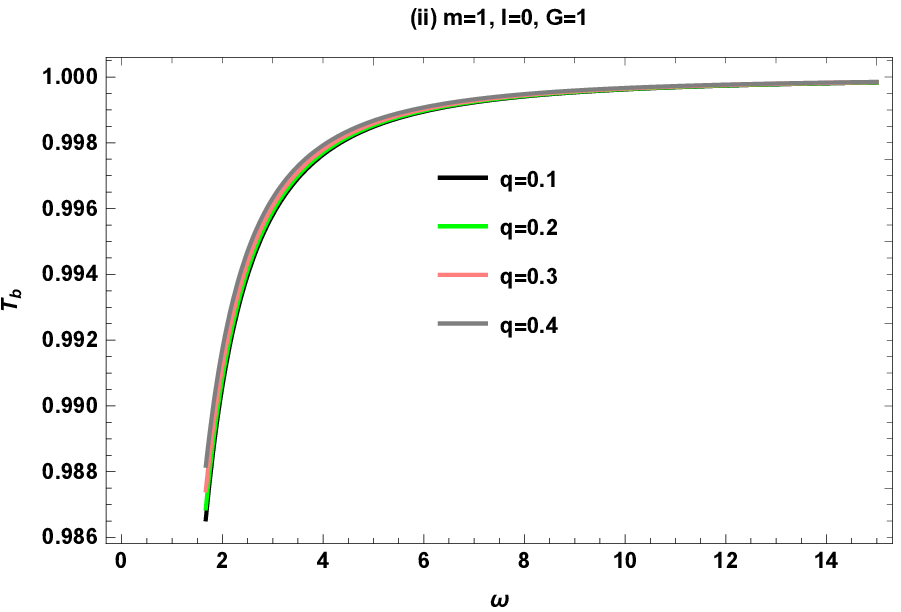,width=0.50\linewidth}\\
{Figure 2: $T_b$ with respect to $\omega$}.
\end{center}
\textbf{Figure 2} shows the behavior of $T_b$ with respect to $\omega$ when mass $m=1$, $l=0$ and $G=1$ are fixed and gives variation to $q$.
In the left figure, it can be seen that the bound $T_b$ is uniformly increasing when domain is taken  $0\leq{\omega}\leq 1$.
In the right figure, it is observe that obtain bound is uniformly negatively decreasing when we taken domain $0\leq{\omega}\leq 15$.\\

\section{Conclusion}

The current paper, is concerning about exploration of deflection angle of MRBH for plasma and non-plasma mediums and greybody bounds. For this purpose, we use GBT and find the deflection angle of photon.\\

\noindent\textbf{1. Bending Angle $\gamma$}\\

\noindent\textbf{(i) Non-plasma Medium}

The deflection angle ${\gamma}$ obtained in Eq.(\ref{S1}) for non-plasma medium depends on $m$ of the BH, $q$, $b$, $G$ and $l$. We observe that in the attain deflection angle the first term is the well known result for the Schwarzschild black hole, when we put $G=1$, $l=0$, $q=0$.\\

\noindent\textbf{(ii) Plasma Medium}

In case of plasma medium, the deflection angle ${\gamma}$ in Eq.(\ref{S2}) depends on $m$ of the BH, $q$, $b$, $G$, $l$ and on the plasma term. The bending angle obtained in the plasma medium increases with the parameter $\frac{\omega_e^2}{\omega_\infty^2}$, which shows that lower the photon frequency observe by a static spectator at infinity, greater the deflection angle of it for the fix electron plasma frequency. We also observe that when we take the $q=0$, $G=1$, $l=0$, the deflection angle obtained in the plasma medium reduces to the deflection angle of the Schwarzschild BH in plasma medium. We also investigate that the deflection angle obtained in the plasma medium reduces to the deflection angle that we have obtained in case of non-plasma, when we take $\frac{\omega_e^2}{\omega_\infty^2}=0$.\\

\noindent\textbf{3. Greybody Factor Bounds $T$}

We calculate the rigorous bound ${T_b}$ in Eq.(\ref{SB19}) for greybody factor. We examined that the obtained bound of greybody factor is depends on the $m$, $G$, $q$, $l$ and $\omega$.\\

\noindent\textbf{2. Graphically}

After computing the bending angles, we also analyzed the graphical behaviour of the bending angles for MRBH in both non-plasma and plasma mediums. We also examined that the graphical behaviour exhibits similar results in both non-plasma and plasma mediums. These results are described as follows:\\

\noindent\textbf{(i) For MRBH }\\

\begin{itemize}
\item\textbf{Bending angle $\gamma$ versus impact parameter $b$}:
The graphical behaviour shows that when we fix the value of $m=1$, $l=0.5$, $G=1$ and gives variation to $q$, the deflection angle shows positive behaviour for $0.6<q<2.09$ and shows negative behaviour at $q=0.6$ and $q=2.09$. Similarly, we observed that deflection angle is decreasing and eventrully goes to infinity for large values of $l$ and by fixing $m=1$, $q=0.8$ and Newton's constant $G=1$.
\end{itemize}

\begin{itemize}
\item\textbf{Bending angle $\gamma$ versus magnetic charge $q$}:
For the fixed values of $G=1$, $m=1$ and $l=0.5$, the $\gamma$ angle is firstly increasing for small values of $b$ but for large values of $b$ deflection angle is decreasing.
\end{itemize}

\textbf{4. Graphical Analysis of Greybody Bound}
\\
At the end, we investigated the graphical results of the greybody bounds for different values of magnetic charges. We have analyzed that bound is increasing when $\omega$ goes from zero to its maximum value at fix $G=1$, $l=0$ and $m=1$.

\end{document}